\documentclass[12pt]{article}
\usepackage{latexsym}
\usepackage{amssymb}
\usepackage{epsf}
\usepackage{amsmath}
\usepackage[hypertex]{hyperref}
\usepackage{graphicx}
\usepackage{color}

\newcommand{\lsim}{\lesssim}
\newcommand{\gsim}{\gtrsim}
\newcommand{\tr}{{\rm Tr}}

\begin{document}
\pagestyle{empty}

\begin{flushright}
LA-UR-09-04505\\
TU-849\\
\end{flushright}

\vspace{3cm}

\begin{center}

{\bf\Large Higgsinoless Supersymmetry and Hidden Gravity}
\\

\vspace*{1.5cm}
{\large 
Michael L. Graesser$^a$, Ryuichiro Kitano$^{b}$ and Masafumi Kurachi$^a$
} \\
\vspace*{0.5cm}

$^a${\it Theoretical Division T-2, Los Alamos National Laboratory, Los
Alamos, NM 87545, USA}\\
$^b${\it Department of Physics, Tohoku University, Sendai 980-8578, Japan}\\
\vspace*{0.5cm}

\end{center}

\vspace*{1.0cm}

\begin{abstract}
{\normalsize
We present a simple formulation of non-linear supersymmetry where
superfields and partnerless fields can coexist. Using this formalism, we
propose a supersymmetric Standard Model without the Higgsino as an
effective model for the TeV-scale supersymmetry breaking scenario. We
also consider an application of the Hidden Local Symmetry in non-linear
supersymmetry, where we can naturally incorporate a spin-two resonance
into the theory in a manifestly supersymmetric way. Possible signatures
at the LHC experiments are discussed.
}
\end{abstract} 

\newpage
\baselineskip=18pt
\setcounter{page}{2}
\pagestyle{plain}
\baselineskip=18pt
\pagestyle{plain}

\setcounter{footnote}{0}

\section{Introduction}

Technicolor is an attractive idea in which electroweak symmetry is
dynamically broken by a strong dynamics operating around the TeV energy
scale~\cite{TC1, TC2}.
The big hierarchy, $m_W \ll M_{\rm Pl}$, is elegantly explained by
the very same reason as $\Lambda_{\rm QCD} \ll M_{\rm Pl}$.
However, it is well-known that there are two phenomenological
difficulties in this idea. One is that the electroweak precision 
measurements seem to prefer scenarios with a weakly coupled 
light Higgs boson~\cite{precision}. 
Another is the difficulty in writing down the Yukawa 
interactions to generate fermion masses in the Standard Model.

After the LEP-I experiments, supersymmetry (SUSY) has become very
popular as a natural scenario for the light weakly coupled Higgs
boson. However, with the experimental bound on the lightest Higgs boson
mass from the LEP-II experiments, parameters in the minimal
SUSY standard model (MSSM) are required to be more and more
fine-tuned, at least in the conventional
scenarios~\cite{Barbieri:2000gf, Kitano:2006gv, Giudice:2006sn}.

In this situation, it may be interesting to (re)consider a hybrid of
technicolor and SUSY along the similar spirit of the early attempts of
SUSY model building~\cite{Dine:1981za,Dimopoulos:1981au}.
We assume that strong dynamics breaks SUSY at the multi--TeV energy
scale (which we call the scale $\Lambda$), with electroweak symmetry
breaking triggered by the dynamics through direct couplings between
the Higgs field and the dynamical sector.
This scenario has several virtues: (1) the Yukawa interactions can be
written down by assuming an existence of elementary Higgs fields in the
UV theory, which mix with (or remain as) the Higgs field to break
electroweak symmetry at low
energy~\cite{Luty:2000fj,Murayama:2003ag,Harnik:2003rs}; (2) the
hierarchy problem, $\Lambda \ll M_{\rm Pl}$, is explained by dynamical
SUSY breaking~\cite{Witten:1981nf}; (3) one can hope that the little hierarchy, 
$m_W \sim m_h \ll \Lambda$, is explained by either SUSY or some other
mechanisms such as the Higgs boson as the pseudo-Nambu-Goldstone
particle in the strong dynamics~\cite{Giudice:2007fh}; (4) the
cosmological gravitino problem is absent~\cite{Viel:2005qj}; (5) one can
expect additional contributions to the Higgs boson mass from the
SUSY breaking sector, with which the mass bound from the LEP-II
experiments can be evaded~\cite{Dine:2007xi}; and (6), there is an
interesting possibility that the LHC experiments can probe the
SUSY breaking dynamics directly. SUSY is phenomenologically motivated from the point (1) (and also (6)) in this framework in addition to the connection to string theory.

Although the TeV-scale SUSY breaking scenario is an interesting
possibility, an explicit model realizing this scenario will not be attempted here.  
In this paper, we take a less ambitious approach and construct an
effective Lagrangian for the scenario without specifying (while hoping
for the existence of) a UV theory responsible for SUSY breaking and its
mediation.

In constructing the effective Lagrangian, we take the following
as organizing principles: (1) the Lagrangian possesses
non-linearly realized supersymmetry;  (2) the quarks/leptons and gauge
fields are only weakly coupled to the SUSY breaking sector, so
that the typical mass splitting between bosons and fermions are
$O(100)$~GeV 
(in other words, the matter and gauge fields are introduced
as superfields which transform linearly under
SUSY); and  
(3) the Higgs boson is introduced as a non-linearly transforming field
because it is assumed to be directly coupled to the SUSY
breaking sector. The Higgsino field is absent in the
minimal model.

In this Higgsinoless model, the Higgs potential receives quadratic
divergences from loop diagrams with the gauge interactions and the Higgs
quartic interaction although the top-quark loops can be cancelled by the
loops of the scalar top quarks as usual. The rough estimate of the
correction to the Higgs boson mass is of the order of $(\alpha/4\pi)
\Lambda^2$ and $(k/16\pi^2) \Lambda^2$ with $k$ being the coupling
constant of the Higgs quartic interaction. By comparing with the
quadratic term needed for electroweak symmetry breaking, $m_H^2 = k
\langle H \rangle^2 / 2$, naturalness suggests $\Lambda \lesssim 4 \pi
\langle H \rangle \sim$ (a few)~$\times$~TeV.  Precision electroweak
constraints, on the other hand, obtained from the LEP-II and SLC
experiments do not generically allow such a low scale without fine-tuning
\cite{Barbieri:1999tm}. The dynamical scale may therefore have to be
larger, $\Lambda \simeq O(6-10 {\rm ~TeV})$. 
To obtain a light Higgs boson at this larger scale either requires
fine-tuning, or some new weakly coupled new physics below $\Lambda$.
(Or simply the Higgsino appears around a few TeV.)
It is also true that the direct coupling to the dynamical sector
generically gives the Higgs boson mass to be $O(\Lambda)$. We may
therefore need to assume that the Higgs boson is somewhat special in the
dynamics, e.g., a pseudo-Goldstone boson.
In this paper we simply ignore the issue because its resolution
depends on the UV completion, and here we only concerned with the
effective theory below the TeV scale. 

The stop potential also receives quadratic divergences, in this case
from a loop diagram involving the Higgs boson (and proportional to
$\lambda_t^2$). This divergence is not cancelled, simply because the
Higgsinos are not present in the low energy theory. One therefore
expects the stops to have a mass no smaller than a loop factor below the
scale $\Lambda$.

The Lagrangian we construct needs to contain interaction terms among
superfields and also partnerless fields such as the Higgs boson. Since
these two kinds of fields are defined on different spaces -- one
superspace and the other the usual Minkowski space -- one needs to
convert the partnerless fields into superfields or vice-versa. 
One approach is to utilize established formulations for constructing
superfields out of partnerless fields~\cite{Rocek:1978nb,Ivanov:1982bpa,Samuel:1982uh}
where the Goldstino field is also promoted to a superfield. In this
paper, we present a simple manifestly supersymmetric formulation where
we do not try to convert partnerless fields into superfields, although
it is totally equivalent to the known formalisms.
The essence is to prepare two kinds of spaces: the superspace and the
Minkowski space, on which superfields and partnerless fields are
defined. By embedding the Minkowski space into the superspace by using a
SUSY invariant map, one can define a Lagrangian density on a single
space-time.
By using the formalism, one can write down a SUSY invariant Lagrangian, in
particular the Yukawa interactions, only with a single Higgs field.
We also find that the coupling constant of the Higgs quartic interaction
can be a free parameter, unrelated to the gauge coupling
constant. Therefore, the Higgs boson mass can be treated as a free
parameter in this model.

As a related topic, a model in which the MSSM is only partly
supersymmetric has been proposed in Ref.~\cite{Gherghetta:2003wm}. There
SUSY is broken explicitly at the Planck scale, and only the Higgs sector
is remained to be supersymmetric which is made possible by a warped
extra-dimension (or a conformal dynamics). Our philosophy is opposite to
that and is, relatively speaking, closer to
Ref.~\cite{Gherghetta:2000qt} by the same authors, where SUSY is broken
on the IR brane (or equivalently by some strong dynamics at the $O({\rm
TeV})$ scale).

As a possible signature of the TeV-scale dynamics, we construct a model
``Hidden Gravity,'' which is an analogy of the Hidden Local Symmetry 
\cite{Bando:1984ej,Bando:1985rf,Bando:1987ym,Bando:1987br,Harada:2003jx} 
in the chiral Lagrangian. The Hidden Local Symmetry is a manifestly chiral
symmetric model to describe the vector resonance (the $\rho$ meson)
as the gauge boson of the hidden vectorial SU(2)
symmetry (the unbroken symmetry of the chiral Lagrangian).
When we apply this technique to SUSY, we obtain a supersymmetric
Lagrangian for a massive spin-two field which is introduced as a
graviton associated with a hidden general covariance because the
unbroken symmetry is the Poincar{\' e} symmetry.

One can consistently incorporate the resonance as a non-strongly coupled
field for a range of parameters and small range of energy.  Indeed, we
show that there is a sensible parameter region where we can perform a
perturbative calculation of the resonant single-graviton production
cross section. At energies not far above the graviton mass the effective
theory becomes strongly coupled and incalculable. If the graviton is
much lighter than the cut-off scale, new physics is required to complete
the theory up to $\Lambda$, another direction not pursued here.  We
discuss signatures of this graviton scenario at the LHC.

\section{Non-linear SUSY and invariant Lagrangian}
\label{sec:nonlinear}

In this section we present a method to construct a Lagrangian invariant
under the non-linearly realized global supersymmetry. We will introduce
the Higgs boson as a non-linearly transforming field (which we call a
non-linear field) and also matter and gauge fields as superfields. We
therefore need a formulation to write down a supersymmetric Lagrangian
where both kinds of fields are interacting.
Ro$\breve{\rm c}$ek~\cite{Rocek:1978nb}, 
 Ivanov and Kapustnikov \cite{Ivanov:1982bpa}, and Samuel and
Wess~\cite{Samuel:1982uh} have established a superfield formalism of
non-linear SUSY by upgrading the Goldstino fermion and other non-linear
fields to constrained superfields. (See \cite{Komargodski:2009rz} for a recent work.)
Although the formalism is somewhat complicated,
using superfields is motivated there as a first step towards embedding
the theory into supergravity.
As we are not interested in supergravity in this paper, we will use a
simpler formalism where the Goldstino field remains as a non-linearly
transforming field. We will also use results from earlier work  by  Ivanov and Kapustnikov 
\cite{Ivanov:1977my} that establishes the correspondence between superfields and non-linear fields \footnote{The generalization of this relationship to local supersymmetry can be found in 
Ref.~\cite{Ivanov:1984hs}.}.

\subsection{Convention and superfields}

We use the metric convention: $\eta_{a b} = {\rm diag.(+---)}$. The
SUSY algebra is
\begin{eqnarray}
 \left \{ 
Q_{\alpha}, \bar Q_{\dot \beta}
\right \}
= 2 \sigma_{\alpha \dot \beta}^{a} P_{a}.
\label{eq:susy}
\end{eqnarray}

Under a group element,
\begin{eqnarray}
g = e^{i c^a P_a + i \eta Q + i \bar \eta \bar Q},
\label{eq:g}
\end{eqnarray}
the superspace coordinate $(x^a, \theta_\alpha, \bar \theta_{\dot
\alpha})$ transforms as~\cite{Salam:1976ib}
\begin{eqnarray}
x^a \to x^{a \prime} =  x^a + c^a + \Delta^a (\eta, \theta),\ \ \ 
\theta_\alpha \to \theta^\prime_\alpha 
= \theta_\alpha + \eta_\alpha,\ \ \ 
\bar \theta_{\dot \alpha} \to 
\bar \theta^\prime_{\dot \alpha} = \bar \theta_{\dot \alpha} + \bar
\eta_{\dot \alpha},
\label{eq:superspace}
\end{eqnarray}
where the $\Delta^a$ factor is defined by
\begin{eqnarray}
 \Delta^a (\eta, \xi) \equiv i \eta \sigma^a \bar \xi
- i \xi \sigma^a \bar \eta .
\end{eqnarray}

A superfield $\Psi (x,\theta,\bar \theta)$ transforms as
\begin{eqnarray}
g \Psi (x,\theta,\bar \theta) g^{-1}
= r(g^{-1}) \Psi (x, \theta, \bar \theta)
= \Psi(x^\prime, \theta^\prime, \bar{\theta}^\prime)
\end{eqnarray}
under SUSY. The operation $r(g^{-1})$ is a representation of
$g^{-1}$ on superfields defined by the second equality.

\subsection{Non-linear SUSY}

The non-linear transformation under $g$ in Eq.~(\ref{eq:g}) is defined
by Volkov and Akulov in Ref.~\cite{Volkov:1972jx}. It is 
\begin{eqnarray}
\tilde x^\mu \to 
\tilde x^{\mu \prime} 
= \tilde x^\mu + c^\mu + \Delta^\mu (\eta, \lambda(\tilde x)),
\label{eq:gc}
\end{eqnarray}
\begin{eqnarray}
\lambda_\alpha (\tilde x) \to \lambda^\prime_\alpha (\tilde x^\prime)
= \lambda_\alpha (\tilde x) + \eta_\alpha,
\label{eq:lam}
\end{eqnarray}
\begin{eqnarray}
\bar \lambda_{\dot \alpha} (\tilde x) \to 
\bar \lambda^\prime_{\dot \alpha} (\tilde x^\prime) 
= \bar \lambda_{\dot \alpha} (\tilde x) + \bar
\eta_{\dot \alpha}.
\label{eq:lam2}
\end{eqnarray}
The fields $\lambda$ and $\bar \lambda$ are the Goldstino fermion and
its complex conjugate, respectively\footnote{Throughout this paper we will 
use the shorthand notation $\lambda=\lambda(\tilde{x})$ with mass 
dimension $-1/2$, unless indicated otherwise.}. 
The transformation above satisfies
the algebra in Eq.~(\ref{eq:susy}).
Note that a global SUSY transformation induces a general
coordinate transformation in Eq.~(\ref{eq:gc}) on the $\tilde x$ space.

One can construct the Maurer-Cartan 1-forms~\cite{Clark:2004xj}:
\begin{eqnarray}
A_\mu^{\ a} = \eta_\mu^{\ a}
- i \lambda \sigma^a \partial_\mu \bar \lambda
+ i \partial_\mu \lambda \sigma^a \bar \lambda,
\label{eq:vielbein}
\end{eqnarray}
\begin{eqnarray}
\nabla_a \lambda = (A^{-1})_a^{\ \mu} \partial_\mu \lambda,
\end{eqnarray}
\begin{eqnarray}
\nabla_a \bar \lambda = (A^{-1})_a^{\ \mu} \partial_\mu \bar \lambda.
\end{eqnarray}
The matrix $A$ transforms as the vielbein under $g$:
\begin{eqnarray}
A_\mu^{\ a} (\tilde x)
\to 
A_\mu^{\prime\ a} (\tilde x^\prime)
=
{\partial \tilde x^\nu \over \partial \tilde x^{\prime \mu}}
A_{\nu}^{\ a} (\tilde x),
\end{eqnarray}
whereas $\nabla_a \lambda$ and $\nabla_a \bar \lambda$ are
invariant.

Matter fields $\phi (\tilde x)$ can be introduced on the $\tilde x$
space. The SUSY transformation on operators is defined by~\cite{Ivanov:1977my}
\begin{eqnarray}
g \phi (\tilde x) g^{-1} 
= \tilde r (h^{-1} (g, \lambda)) \phi (\tilde x)
= \phi (\tilde x^\prime),
\end{eqnarray}
where $\tilde r(h^{-1})$ is the representation of the space-time
translation acting on the $\tilde x$ space defined in
Eq.~(\ref{eq:gc})\footnote{In terms of the same notation in
Eqs.~(\ref{eq:lam}) and (\ref{eq:lam2}), the transformation of the
classical field (or expectation values of the field) is: $\phi(\tilde x)
\to \phi^\prime (\tilde x^\prime) = \phi (\tilde x)$. For a classical
superfield, $\Psi(x,\theta,\bar \theta) \to \Psi^\prime (x^\prime,
\theta^\prime, \bar \theta^\prime) = \Psi (x, \theta, \bar \theta)$. The
transformation laws remain unchanged for fields with Lorentz indices.}.
A supersymmetric action for $\phi(\tilde x)$ can be obtained simply by
writing an invariant action under the general coordinate transformation
in Eq.~(\ref{eq:gc}) by using the vielbein in Eq.~(\ref{eq:vielbein}).

Superfields and non-linear fields are living in different spaces $x$ and
$\tilde x$ which we cannot identify as the same space at this stage
since their SUSY transformations are different. In order to
write down an interaction term between $\phi(\tilde x)$ and superfields,
we need a ``converter'' which transforms a field in the $\tilde x$ space
into a superfield in the superspace $(x,\theta, \bar \theta)$.

The discussion is completely parallel to the formalism of
Callan-Coleman-Wess-Zumino (CCWZ) for internal global
symmetries~\cite{CCWZ}. (See also~\cite{Bando:1987br} for a
review.) There a global symmetry $G$ is spontaneously broken down to a
subgroup $H$. A Lagrangian which is invariant under the global $H$
transformation can be upgraded to a $G$ invariant one by making the
Lagrangian invariant under a local $H$ transformation where the
Maurer-Cartan 1-form (projected onto the unbroken generators) can be
used as the gauge connection.

In the CCWZ formalism, a linear representation of a group element
$\xi(x) \equiv e^{i \pi(x)} \in G$ plays a role of the converter between
the $G$ and $H$ indices by defining the transformation of $\xi$ to be
$\xi(x) \to g \xi (x) h^{-1} (g, \pi)$.
We can follow a similar prescription here by taking the converter
$r(\Xi)$ with $\Xi = e^{iQ\lambda + i\bar Q \bar \lambda}$.
For example, a superfield $\Phi (x, \theta, \bar \theta)$ can be
constructed from a non-linear field $\phi(\tilde x)$ by
\begin{eqnarray}
\Phi (x, \theta, \bar \theta) \equiv 
r(\Xi) \phi(x) =
\phi(x - \Delta(\lambda (\tilde x),\theta)).
\end{eqnarray}
At this stage, $\Phi$ is not defined yet because the last expression
still contains $\tilde x$. 
The appropriate identification is found to be
\begin{eqnarray}
\tilde x^\mu = x^\mu - \Delta^\mu (\lambda (\tilde x), \theta) ,
\label{eq:convert}
\end{eqnarray}
which consistently defines the superfield
$\Phi$~\cite{Ivanov:1977my}. (See also
\cite{Rocek:1978nb,Samuel:1982uh,Uematsu:1981rj,Samuel:1983jp,Luty:1998np,Antoniadis:2004uk} for
constructing superfields out of non-linear fields).
There is still a little bit of complication because the above equation
is non-linear in $\tilde x$.
It is possible to iteratively solve $\tilde x$ in terms of $x$,
$\lambda(x)$ and $\theta$, but the solution involves many terms although
the iterations will be terminated at finite steps.
Nonetheless, one can explicitly check that $\Phi (x, \theta, \bar
\theta)$ is a superfield, i.e.,
\begin{eqnarray}
\Phi (x^\prime, \theta^\prime, \bar \theta^\prime) = \phi(\tilde x^\prime),
\end{eqnarray}
because
\begin{eqnarray}
\tilde x^{\prime \mu} =  x^{\prime \mu} - \Delta^\mu (\lambda^\prime
 (\tilde x^\prime), \theta^\prime ).
\end{eqnarray}
In general, any function of
\begin{eqnarray}
\phi(\tilde x),\ \ \  \theta - \lambda (\tilde x),
\ \ \  \bar \theta - \bar \lambda (\tilde x),
\end{eqnarray}
with $\tilde x$ defined by Eq.~(\ref{eq:convert}) is a superfield.

As an equivalent formulation, one can construct a supersymmetric action
using the supersymmetric invariant 
\begin{eqnarray}
1 = \int d^4 \tilde x \det X 
\delta^4 (x^\mu - \tilde x^\mu - \Delta^\mu (\lambda (\tilde x) , \theta )
),
\end{eqnarray}
in the superspace integral. The Jacobian matrix $X$ is 
\begin{eqnarray}
X_\mu^{\ a} = \eta_\mu^{\ a} 
- i\theta \sigma^a \partial_\mu \bar \lambda
+ i \partial_\mu \lambda \sigma^a \bar \theta,
\end{eqnarray}
which transforms in the same way as $A$, and is equal to $A$ at
$\theta=\lambda(\tilde{x})$, $\bar{\theta} =
{\bar \lambda}(\tilde{x})$.
With the delta function, one can treat $x$, $\theta$, $\bar \theta$, and
$\tilde x$ as independent variables in constructing the Lagrangian.
The invariant action can be written down as
\begin{eqnarray}
S &=& \int d^4 x d^4 \theta d^4 \tilde x \det X \
\delta^4 (x^\mu - \tilde x^\mu - \Delta^\mu (\lambda (\tilde x) , \theta ) )
\nonumber \\
&&
\times {\cal K} \left[
\Psi (x, \theta, \bar \theta),
\phi (\tilde x),
\theta - \lambda,
\bar \theta - \bar \lambda,
\nabla_a \lambda,
\nabla_a \bar \lambda,
A, X, \cdots
\right],
\end{eqnarray}
where $\Psi$ and $\phi$ represents arbitrary superfields and non-linear
fields, respectively. The function ${\cal K}$ must be real and scalar
under the general coordinate transformation about the $\tilde x$
coordinate.
As an example of the invariant action, we can take ${\cal K} = \Psi (x,
\theta, \bar \theta)$ which gives the supersymmetric action, 
\begin{eqnarray}
S = \int d^4 x d^4 \theta \Psi (x, \theta, \bar \theta).
\end{eqnarray}
If we take ${\cal K} = \delta^4 (\theta - \lambda) \cdot (-f^4/2)$, we
obtain the Volkov-Akulov action~\cite{Volkov:1972jx}
\begin{eqnarray}
S = -{f^4 \over 2} \int d^4 \tilde x \det A.
\end{eqnarray}
This contains the kinetic term for the Goldstino. The parameter $f$ is
the decay constant which represents the size of the SUSY breaking. Note
that naive dimensional analysis \cite{Manohar:1983md} implies a cutoff
$\Lambda \sim \sqrt{4 \pi} f$\footnote{Although the Volkov-Akulov action
involves many terms with different numbers of derivatives, the momentum
expansion still makes sense once we fix the number of external lines in
each amplitude. For example, the lowest order (tree) amplitudes with $d$
external Goldstinos is $O(p^d)$ and $n$-loop corrections to that are
$O(p^{d+4n})$. When comparing with other terms in the action, we should
count the number of derivatives with a fixed number of the Goldstino
fields. We can easily see that terms in Eq.~(\ref{eq:VA2}) and
(\ref{eq:R}) contain more derivatives than the Volkov-Akulov
action.}.

One may generalize the Volkov-Akulov action. From the invariance of
$\nabla_a \lambda$,
\begin{eqnarray}
S = -{f^4 \over 2} \int d^4 \tilde x \det A 
~F(\nabla_a \lambda, \nabla_b {\bar \lambda})
\label{eq:VA2}
\end{eqnarray} 
is SUSY invariant for any $F$ that forms a Lorentz invariant out of
$\nabla_a \lambda$ and/or $\nabla_b {\bar
\lambda}$~\cite{Uematsu:1981rj, Clark:2004xj}. Another possibility is to
consider the ``metric''
\begin{eqnarray} 
G_{\mu \nu} \equiv A_\mu^{\ a} A_{\nu}^{\ b} \eta_{ab}
\end{eqnarray} 
which transforms as a covariant tensor and can be used to build
invariant actions. For instance,
\begin{eqnarray} 
 \int d^4 \tilde x \det A ~R[G]
\label{eq:R}
 \end{eqnarray} 
is invariant under global SUSY transformations. The leading term begins
at $O(\partial^3)$ and involves two Goldstinos. Terms involving four
Goldstinos have $O(\partial^4)$ and so on, with the last term involving 8
Goldstinos and 6 derivatives.

Lagrangian densities with a single space-time coordinate $x$ or $\tilde
x$ are obtained by performing one of the space-time integrals, i.e., of
the form:
\begin{eqnarray}
S = \int d^4 x {\cal L}(x) + \int d^4 \tilde x \tilde {\cal L}(\tilde x) ~.
\end{eqnarray}
This action is of course identical to
\begin{eqnarray}
S= \int d^4 x \left(
{\cal L}(x) + \tilde {\cal L} (x)
\right).
\end{eqnarray}
We now have a Lagrangian in a single space-time.

Superpotential-like terms can also be constructed as
\begin{eqnarray}
S &=& \int d^4 y d^2 \theta d^4 \tilde x \det Y
\delta^4 (y^\mu - \tilde x^\mu - i \lambda \sigma^\mu \bar \lambda
+ 2 i \theta \sigma^\mu \bar \lambda)
\nonumber \\
&& \times {\cal W}
\left[
\Psi (y, \theta),
\phi(\tilde x),
\theta - \lambda,
\nabla_a \lambda,
\nabla_a \bar \lambda,
A, \cdots
\right] + {\rm h.c.},
\end{eqnarray}
with
\begin{eqnarray}
y^\mu = x^\mu - i \theta \sigma^\mu \bar \theta,
\end{eqnarray}
and
\begin{eqnarray}
Y_\mu^{\ a} = \eta_\mu^{\ a}
+ i \partial_\mu \lambda \sigma^a \bar \lambda
+ i \lambda \sigma^a \partial_\mu \bar \lambda
- 2 i \theta \sigma^a \partial_\mu \bar \lambda.
\end{eqnarray}

There is an intuitive picture for this construction. We can imagine a
set-up where a 3-brane is embedded into a superspace. The Goldstino
field $\lambda(\tilde x)$ defines the map from a point on the brane to a
point in the superspace.
One can write down a usual superspace Lagrangian as well as a brane
localized action. The brane action should be invariant under the general
coordinate transformation because SUSY, which is a translation
in the superspace, induces a coordinate transformation (which depends on
$\lambda(\tilde x)$) on the brane.
The interaction terms between superfields (bulk fields) and brane fields
can be written down by using a delta function.

\subsection{Gauge invariance}

It is now possible to write down an interaction term between a chiral
superfield ${\cal O} (y, \theta)$ and a non-linear field $\phi (\tilde
x)$:
\begin{eqnarray}
S = \int d^4 \tilde x d^2 \theta \det A \ 
{\cal O}(\tilde x^\mu + i \lambda \sigma^\mu \bar \lambda - 2i \theta \sigma^\mu
\bar \lambda, \theta) \phi(\tilde x) + {\rm h.c.},
\label{eq:mix}
\end{eqnarray}
from
\begin{eqnarray}
{\cal W} = {\cal O} (y, \theta) \phi(\tilde x).
\end{eqnarray}
By taking ${\cal O}$ as a bilinear of the quarks/leptons superfields and
$\phi$ as the Higgs boson, this gives a supersymmetric Yukawa
interaction term.

However, if ${\cal O}$ and $\phi$ are charged under some gauge symmetry
(as it is true in the Standard Model), we need to modify the
interaction term since it is not gauge invariant.
The gauge transformation is defined by
\begin{eqnarray}
{\cal O} \to e^{i \Lambda^a (y, \theta) \tilde T^a} {\cal O} (y, \theta),
\end{eqnarray}
and 
\begin{eqnarray}
\phi \to e^{i \alpha^a (\tilde x) T^a} \phi (\tilde x),
\end{eqnarray}
where $(\tilde T^a)_{ij} = - (T^a)_{ji}$.
Under this transformation, the action is clearly not invariant.

In order to maintain gauge invariance, we write the action as
\begin{eqnarray}
S &=& \int d^4 x d^4 \theta d^4 \tilde x  \ \det X \
\delta^4 (x - \tilde x - \Delta (\lambda, \theta))\ 
\delta^4 (\theta - \lambda)
\left( {1 \over 2}
e^{2gV} D^2 e^{-2gV}
{\cal O} \right)
\phi(\tilde x)
\nonumber \\
&=&
\int d^4 \tilde x \det A \ \left.
\left(
{1 \over 2}e^{2gV} D^2 e^{-2gV}
{\cal O} \right) \right|_{x = \tilde x, 
\theta = \lambda, 
\bar \theta = \bar \lambda}
\phi(\tilde x),
\label{eq:yukawa}
\end{eqnarray}
where $V = V^a \tilde T^a$. The gauge transformation of the vector
superfield is
\begin{eqnarray}
e^{-2gV} \to e^{i \Lambda^\dagger} e^{-2gV} e^{-i \Lambda}.
\end{eqnarray}
The derivative operators are defined by
\begin{eqnarray}
D_{ \alpha} = {\partial \over \partial \theta^{\alpha}}
- i (\sigma^a \bar \theta)_{\alpha} {\partial \over \partial
x^a},\ \ \ 
\bar D_{\dot \alpha} = - {\partial \over \partial \bar \theta^{\dot \alpha}}
+ i (\theta \sigma^a )_{\dot \alpha} {\partial \over \partial x^a} .
\end{eqnarray}
By defining the gauge transformation of $\phi(\tilde x)$ with
\begin{eqnarray}
\alpha^a(\tilde x) =\Lambda^a(y,\lambda)= \Lambda^a(\tilde x, \lambda, \bar \lambda),
\label{alpha gauge transformation}
\end{eqnarray}
the interaction term in Eq.~(\ref{eq:yukawa}) is gauge invariant.
Note, however, that the function $\alpha(\tilde x)$ defined above can be
complex valued unlike the usual gauge transformation (it is real in Wess-Zumino gauge).

It is possible to define a covariant derivative to write down a kinetic
term for $\phi(\tilde x)$. We define gauge
superfields~\cite{Bailin:1994qt}
\begin{eqnarray}
g {\cal A}_a (x, \theta, \bar \theta) \equiv 
{1 \over 4} \bar D e^{2 g V} \bar \sigma_a D e^{-2 g V},
\end{eqnarray}
and
\begin{eqnarray}
g {\cal A}_\alpha (x, \theta, \bar \theta) \equiv
e^{2 g V} D_\alpha e^{-2 g V}.
\end{eqnarray}
The gauge transformations of these superfields are
\begin{eqnarray}
{\cal A}_a
\to e^{i \Lambda} {\cal A}_a e^{-i\Lambda}
+ {i \over g} e^{i \Lambda} {\partial \over \partial x^a} e^{-i \Lambda},
\end{eqnarray}
\begin{eqnarray}
{\cal A}_\alpha
\to e^{i \Lambda} {\cal A}_\alpha e^{-i \Lambda}
+ {1 \over g} e^{i \Lambda} D_\alpha e^{-i \Lambda}.
\end{eqnarray}
We defined $V = V^a T^a$ and $\Lambda = \Lambda^a T^a$ this time.  By
using these superfields, the ``covariant'' derivative is constructed as
\begin{eqnarray}
D_a \equiv \nabla_a - i g {\cal A}_a
+ g (\nabla_a \lambda^\alpha ) {\cal A}_\alpha.
\label{eq:cov-der}
\end{eqnarray}
This derivative operator is not covariant under the gauge transformation
at this stage. However, under the $\delta$-functions, $\delta^4 (x -
\tilde x - \Delta)$ and $\delta^4(\theta - \lambda)$, one can confirm
that it behaves as a covariant derivative: $D_a \phi(\tilde x) \to e^{i
\alpha} D_a \phi(\tilde x)$.

Then the kinetic term can be written as
\begin{eqnarray}
{\cal K}_{\rm kin.} = 
\delta^4 (\theta - \lambda)
\left[
(D_a \phi (\tilde x))^\dagger e^{-2 g V} D^a \phi (\tilde x)
\right].
\end{eqnarray}
The potential terms are
\begin{eqnarray}
{\cal K}_{\rm pot.} = \delta^4 (\theta - \lambda)
\left[
- m^2 \phi^\dagger (\tilde x) e^{-2 g V} \phi (\tilde x) 
- {k \over 4} \left( \phi^\dagger (\tilde x) 
e^{-2 g V} \phi (\tilde x) \right)^2
\right].
\label{eq:pot}
\end{eqnarray}
Both are gauge invariant under the delta functions with complex-valued
$\alpha$ satisfying Eq. (\ref{alpha gauge transformation}).
The quartic coupling $k$ is unrelated to the gauge coupling constant in
contrast to the prediction of the MSSM.

\section{Higgsinoless SUSY}
\label{sec:higgs}

We are now ready to construct a Lagrangian. For quarks/leptons and gauge
superfields, one can simply write down the MSSM Lagrangian in the
superspace.
Soft SUSY breaking terms can be written down by using delta
functions $\delta^2 (\theta - \lambda)$ and $\delta^4 (\theta -
\lambda)$:
\begin{eqnarray}
&& {\cal K}_{\rm soft} = - \delta^4 (\theta - \lambda) \cdot 
m_{\Psi}^2 \Psi^\dagger
 \Psi
\nonumber \\
&\Rightarrow&
S \ni - \int d^4 \tilde x \det A\  m_\Psi^2 \Psi^\dagger \Psi (\tilde x,
\lambda, \bar \lambda),
\end{eqnarray}
and 
\begin{eqnarray}
&& {\cal W}_{\rm soft} = - \delta^2 (\theta - \lambda) 
\cdot {m_{1/2} \over 2} W^\alpha W_\alpha
\nonumber \\
&\Rightarrow&
S \ni - \int d^4 \tilde x
\det A \ 
{m_{1/2} \over 2} W^\alpha W_\alpha (\tilde x, \lambda, \bar \lambda) +
{\rm h.c.}
\end{eqnarray}
These are the same as the spurion method for the soft SUSY
breaking terms. The appearance of the Goldstino interactions makes these
terms manifestly supersymmetric.
One can also add hard breaking terms by using covariant derivatives. We
assume that such soft and hard breaking terms are somewhat suppressed
because the quarks/leptons and gauge fields are not participating the
SUSY breaking dynamics.

We introduce the Higgs field as a non-linear field on the $\tilde x$
space, $h(\tilde x)$, motivated by an assumption that the SUSY
breaking dynamics at the cut-off scale $\Lambda$ has something to do
with the origin of electroweak symmetry breaking.
The way to construct interaction terms has been discussed already in the
previous subsection.
The Yukawa interactions for up-type quarks are
\begin{eqnarray}
{\cal K}_{\rm up} = \delta^4 (\theta - \lambda) 
\left[
y_u^{ij} h(\tilde x) \cdot 
\left(
{1 \over 2}D_{\rm (cov)}^2 U^c_j Q_i 
\right)
\right]
.
\end{eqnarray}
For down-type quarks and leptons, 
\begin{eqnarray}
{\cal K}_{\rm down} &=& \delta^4 (\theta - \lambda)
\nonumber \\
&&
\times
\left[
y_d^{ij} h(\tilde x)^\dagger e^{-2 g V} 
\left(
{1\over 2}D_{\rm (cov)}^2 D^c_j Q_i 
\right)
+ y_e^{ij} 
h(\tilde x)^\dagger e^{-2 g V} 
\left(
{1\over 2}D_{\rm (cov)}^2 E^c_j L_i 
\right)
\right].
\nonumber \\
\end{eqnarray}
Here we have used the covariant derivative:
\begin{eqnarray}
D_{\rm (cov)}^2 \equiv e^{2 g V} D^2 e^{-2 g V}.
\end{eqnarray}
It is not necessary to introduce two kinds of Higgs fields for the
Yukawa interactions. 
The $A$-terms can also be written down by taking
\begin{eqnarray}
{\cal W}_{\rm A} = \delta^2 (\theta - \lambda ) 
\left[
A_u^{ij}  h(\tilde x) \cdot ( U^c_j Q_i )
\right],
\end{eqnarray}
and
\begin{eqnarray}
 {\cal K}_{\rm A} &=& \delta^4 (\theta - \lambda)
\left[
A_d^{ij} h(\tilde x)^\dagger e^{-2 g V} 
(D^c_j Q_i )
+ A_e^{ij} h(\tilde x)^\dagger e^{-2 g V}
(E^c_j L_i )
\right].
\nonumber \\
\end{eqnarray}

Since the quartic coupling of the Higgs boson in Eq.~(\ref{eq:pot}) is a
free parameter, the Higgs boson mass is not related to the $Z$-boson
mass. It is not a very obvious result that we could write down a
Lagrangian with a single Higgs boson with the enlarged gauge
invariance. For example, in Ref.~\cite{Samuel:1983jp} it has been
necessary to introduce an extra Higgs boson, and that is claimed to be a
general requirement for constructing a realistic model with non-linear
SUSY.


\section{Hidden Gravity}
\label{sec:gravity}

A SUSY transformation in the $\tilde x$ space is realized as a local
coordinate transformation in Eq.~(\ref{eq:gc}). This local translation
allows us to introduce a metric in the $\tilde x$ space having a local
transformation law under the global SUSY. This provides a description of
a composite spin-two field\footnote{An attempt to describe a spin-two
resonance in QCD as a massive graviton can be found in
Ref.~\cite{Isham:1971gm}, whose supergravity extension is
discussed in Ref.~\cite{Chamseddine:1978yu}. A more ambitious attempt to
formulate Einstein gravity as a composite of the Goldstino fermions can
be found in Ref.~\cite{Lukierski:1982vw}. The appearance of a massive bound-state graviton in open string field theory can be found in Ref. \cite{Siegel:1993sk}.} in the SUSY breaking dynamics
analogous to the $\rho$ meson in QCD. We further elaborate on this
comparison towards the end of this section.

Specifically, we introduce a second ``metric'' whose transformation 
under $g$ is
\begin{eqnarray}
g_{\mu \nu} (\tilde x) \to g^\prime_{\mu \nu} (\tilde x^\prime) 
= 
{\partial \tilde x^\rho \over \partial \tilde x^{\prime \mu}}
{\partial \tilde x^\sigma \over \partial \tilde x^{\prime \nu}} g_{\rho
\sigma} (\tilde x),
\end{eqnarray}
where $\tilde x^\prime$ is given in Eq.~(\ref{eq:gc})\footnote{We could
have instead defined the spin-two field to transform as a scalar under
$g$: $g_{ab}(\tilde{x}) \to g^\prime_{ab} (\tilde x^\prime) =
g_{ab}(\tilde{x})$. This is an equivalent formulation, since the two
definitions are related by multiplying by the vielbein,
$g_{ab}(\tilde{x}) \equiv A^{\mu}_{~a} A^{\nu}_{~b} ~g_{\mu
\nu}(\tilde{x}) $. We will not pursue this formulation any further.}.
Note that this is a {\it global} SUSY transformation, and one should not
be confused with the actual general coordinate transformation on the
$x$-space. The space-time is always flat. The deviation of $g_{\mu \nu}$
from the Minkowski metric describes the spin-two field.

The invariant action having the Fierz-Pauli form~\cite{Fierz:1939ix} is
\begin{eqnarray}
S = \int d^4 \tilde x
\left[
-{f^4 \over 2} \det A
- {m_{\rm P}^2 \over 2} \sqrt g R(g)
- {m_{\rm P}^2 m^2 \over 8} \sqrt g
g^{\mu \nu} g^{\alpha \beta} \left(
H_{\mu \alpha} H_{\nu \beta} - H_{\mu \nu} H_{\alpha \beta}
\right)
\right]
\label{eq:action}
\end{eqnarray}
where
\begin{eqnarray}
H_{\mu \nu} = g_{\mu \nu} - G_{\mu \nu}
\end{eqnarray} 
and  
\begin{eqnarray} 
G_{\mu \nu} = A_\mu^{\ a} A_{\nu}^{\ b} \eta_{ab}.
\end{eqnarray} 
is a covariant tensor, defined previously.  The $H_{\mu \nu}$ field is
therefore a SUSY covariant tensor. The scale $m_{\rm P} $ is a mass
parameter of $O({\rm TeV})$, unrelated to the four-dimensional Planck
mass $G^{-1/2}_N$ of Einstein gravity.  With
\begin{eqnarray} 
g_{\mu \nu} = \eta_{\mu \nu} + \frac{2}{m_{\rm P}} h_{\mu \nu}~,
\end{eqnarray}
one has 
\begin{eqnarray} 
H_{\mu \nu} =  \frac{2}{m_{\rm P}}  h_{\mu \nu} + 
\left(i \lambda \sigma_{\mu} \partial_{\nu} {\bar \lambda} 
- i \partial_{\nu} \lambda \sigma_{\mu} {\bar \lambda} + (\mu
\leftrightarrow \nu) \right)+ \mbox{(four-fermion terms)}
\end{eqnarray}
Note the relative coefficient (of $-1$) between the two terms appearing
in the definition of $H_{\mu \nu}$ is fixed by requiring that the
Fierz-Pauli mass term not introduce a tadpole for the graviton.  The
last term in the action gives a mass $m$ to the spin-two field in a
global SUSY invariant way.

There are other invariant terms involving $h_{\mu \nu}$ and up to two
derivatives, such as
\begin{eqnarray}
\sqrt g, ~\det A \cdot R(g), \ \ \mbox{etc}.  
\end{eqnarray}
but these are forbidden by the Lorentz invariance of the vacuum and the
absence of ghosts and tachyons. That is, the Einstein action with
Fierz-Pauli mass term is the unique tachyon and ghost-free action for a
spin-two field~\cite{vanDam:1970vg, VanNieuwenhuizen:1973fi}.  Although
loop corrections will not preserve this form, the ghost pole is harmless
since its effect is pushed to the cutoff \cite{ArkaniHamed:2002sp}.

Other interactions, such as 
\begin{eqnarray} 
\sqrt{g} \cdot R(H),~\det A \cdot R(H)
\end{eqnarray}
begin at higher than quadratic order in $h_{\mu \nu}$. 

In the chiral Lagrangian of QCD one can construct SU(2) vector- and
axial-type 1-forms $j_{V,A}$ out of the pion fields. A chirally
invariant Lagrangian can be constructed only out of $j_A$, since $j_V$
transforms inhomogeneously under the chiral SU(2).
By introducing an SU(2)$_V$ vector boson $V_{\mu}$, the term ${\rm Tr}
\tilde{j}_V \tilde{j}_V$, with $\tilde{j}_V = j_V - V$, is made chirally
invariant and can be added to the action. This gives a mass to the
vector boson, but no kinetic term; it can be trivially integrated
out. The key assumption
of~\cite{Bando:1984ej,Bando:1985rf,Bando:1987ym,Bando:1987br,Harada:2003jx}
is that this vector boson is dynamical (and describes the $\rho$ vector
meson).  The action obtained in this way coincides with the
spontaneously broken SU(2)$_V$ gauge theory in the unitary gauge, which
obviously can have a sensible description up to some high energy scale.

The analogy of the massive spin-two field as formulated here to the
Hidden Local Symmetry (HLS) of QCD can now be drawn more closely, though
imprecisely. Here the Maurer-Cartan 1-forms $A_{\mu} ^{~a}$ are
analogous to the $j_V$ in QCD. By introducing a spin-two field having a
local and inhomogeneous transformation under the global symmetry, it is
then possible to introduce the 1-forms into the action, in the form of
the Fierz-Pauli mass term\footnote{The other invariant is $\det A$,
since the coordinates also transform.}. Further assuming a Ricci scalar
term in the action for the spin-two field is equivalent to the physical
assumption in HLS that the gauge boson is dynamical.


\subsection{Perturbative Unitarity} 
\label{graviton unitarity bounds} 

A question to be addressed here is whether this new spin-two resonance
can be consistently introduced in a weakly coupled regime, in which
perturbative calculations make sense at energies of $O(m)$.  We first check this by looking at
the elastic scattering of two Goldstinos with the same helicity. Then we
require that the spin-two field is not strongly coupled at threshold.

In supergravity, the amplitudes of the elastic scattering of two
gravitinos have been calculated in Ref.~\cite{Casalbuoni:1988sx} and it
has been shown that the scalar partners of the Goldsino fermion
unitarize the amplitudes if they are light enough. We show in this
subsection that a spin-two field, instead of the scalar fields, can
also partially cancel the growth of the scattering amplitudes.
This is analogous to the discussion of the $WW$ scattering in the
Standard Model. The SU(2)$_L$ partner of the Goldstone boson, the Higgs
boson, can unitarize the $WW$ scattering amplitude if it is light
enough. But alternatively, it has been known from the analysis of the
Higgsless model~\cite{Higgsless1,Higgsless2} that a massive vector boson
can also partially cancel the amplitude, and the theory can remain
perturbative up to some high energy scale above the Kaluza-Klein
scale~\cite{Higgsless1,uni}.  The massive vector boson is indeed
identified with the one in HLS once the Higgsless model is formulated as
a four-dimensional theory~\cite{decHless1, decHless2, Chivukula:2006cg,
BESS} by using the technique of deconstruction~\cite{deconstruction}.

The amplitude $\cal{M}_{\lambda \lambda}$ for $\lambda \lambda \to
\lambda \lambda$ receives contributions from both the Volkov-Akulov
action and the action for the spin-two field.  Specifically,
one obtains  
\begin{eqnarray}
{\cal M}_{\lambda \lambda} &=& 
{\cal M}_{\lambda \lambda}^{(\det A)} 
+ {\cal M}_{\lambda \lambda}^{(HH)}, \label{eq:Mlamlam}\\
{\cal M}_{\lambda \lambda}^{(\det A)} &=& \frac{2 s^2}{f^4}, \label{eq:Mdet}\\
{\cal M}_{\lambda \lambda}^{(HH)} &=& -\frac{5 m_{\rm P}^2 m^2 s^2}{f^8} \nonumber \\
 & & -\frac{m_{\rm P}^2 m^2 s^2}{f^8}
 \left[ 
 \frac{m^2}{t-m^2}\left(\frac{5}{2}+\frac{3}{2}\cos \theta \right)  + 
 \frac{m^2}{u-m^2}\left(\frac{5}{2}-\frac{3}{2}\cos \theta \right)
 \right]. \label{eq:MHH}
\end{eqnarray}
The contribution from the Volkov-Akulov action is given in
Eq.~(\ref{eq:Mdet}), and those from the spin-two action in
Eq.~(\ref{eq:MHH}). 
The production angle $\theta$ is defined in the center-of-mass frame.

The contributions from the spin-two action deserve further comment. The
second term in the RHS of Eq.~(\ref{eq:MHH}) is the contribution from
$t$ and $u$ channel exchange of the massive graviton, arising from the
Goldstino-Goldstino-graviton coupling in the Fierz-Pauli mass term.  The
Fierz-Pauli mass term however also has a contact four-point interaction
involving the Goldstinos, giving the first term in the RHS of
Eq.~(\ref{eq:MHH}).
By inspection, in the low-energy limit these two contributions to ${\cal
M}_{\lambda \lambda}^{(HH)}$ exactly cancel. That is, at low energies
one obtains the same $\lambda \lambda$ scattering amplitude as in the
theory without a massive graviton. 
Therefore, the decay constant $f$ appearing in the action in
Eq.~(\ref{eq:action}) is the same as the one in the original
Volkov-Akulov action.

\begin{figure}[t]
\begin{center}
\includegraphics[height=6cm]{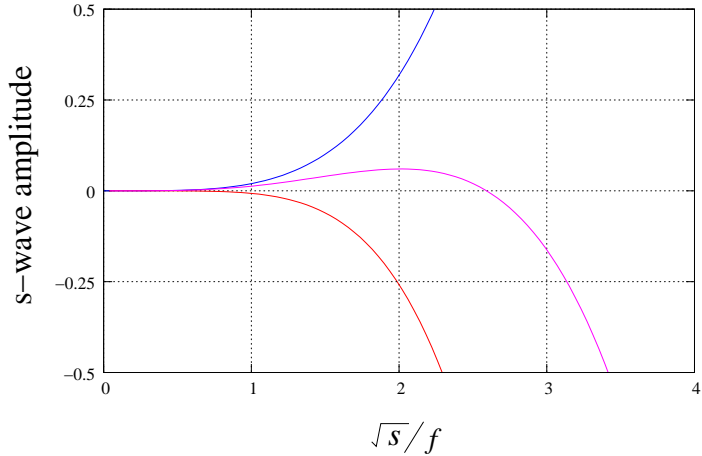}
\end{center}
\caption{The magnitude of the $s$-wave amplitude 
as a function of $\sqrt{s}/f$. Upper 
and lower curves represent contributions from $(-{f^4/2}) \det A$ and 
$(H_{\mu\nu}H^{\mu\nu}-H_\mu^\mu H_\nu^\nu)$ terms, and 
the middle curve represents the total amplitude.
Here, $m_{\rm P}=0.7f$ and $m=1.2f$ are used as an example.
}
\label{fig:partial}
\end{figure}

The partial wave amplitudes are defined by  
\begin{equation}
 {\cal M}_{\lambda \lambda}^{\ell} 
 = \frac{1}{64\pi} \int_{-1}^{1} d(\cos \theta) P_\ell(\cos \theta) 
   {\cal M}_{\lambda \lambda}, 
\end{equation}
where $P_\ell$ denotes the Legendre Polynomials: $P_0(z)=1$, $P_1(z)=z$,
$P_2(z)=\frac{1}{2}(3z^2-1)$, etc. Since the particles in the final
state are identical, we compensate the integral over all of phase space
by multiplying by a factor of $1/2$.  Substituting the expressions in
Eqs.~(\ref{eq:Mlamlam}--\ref{eq:MHH}), we obtain the following
$s$-wave amplitude for the $\lambda \lambda$ scattering.
\begin{eqnarray}
 {\cal M}_{\lambda \lambda}^{\ell=0}
 &=&  \frac{1}{16\pi}\frac{s^2}{f^4}  \nonumber \\
 && \!\!\!\!\! -\frac{5}{32\pi} \frac{m_{\rm P}^2 m^2 s^2}{f^8} 
 - \frac{1}{16\pi}\frac{m_{\rm P}^2 m^4 s}{f^8}  
 \left[ 
 3 - \left( 4+\frac{3m^2}{s} \right) \ln \left( 1+\frac{s}{m^2} \right)
 \right]. \label{eq:partial}
\end{eqnarray}
The first term in the RHS comes from ${\cal M}_{\lambda \lambda}^{(\det
A)}$, while the remaining terms come from ${\cal M}_{\lambda
\lambda}^{(HH)}$.  
There is no parameter to control the relative sign of the two
contributions.  One can see that two $O(s^2)$ terms in
Eq.~(\ref{eq:partial}) always have an opposite sign, and thus the
graviton contribution partially cancels the growth of the amplitude.
The magnitude of the $s$-wave amplitude is plotted in
Fig.~\ref{fig:partial} as a function of $\sqrt{s}/f$.
The upper and lower curves represent contributions from $(-f^4/2) \det
A$ and $(H_{\mu\nu}H^{\mu\nu}-H_\mu^\mu H_\nu^\nu)$ interactions (i.e.,
the first and second row of the RHS of Eq.~(\ref{eq:partial}),
respectively). The the middle curve represents the total amplitude. The
parameters $m_{\rm P}=0.7f$ and $m=1.2f$ are chosen for illustration.

We define the perturbative-unitarity-violation scale $E_\ast$, by the
energy where the tree level $s$-wave amplitude of $\lambda \lambda$
scattering reaches the value $0.5$. In the case of the example depicted
in Fig.~\ref{fig:partial}, the pure Goldstino amplitude (i.e., the upper
curve) gives $E_\ast \sim 2.2f$. The contribution of the spin-two
particle to this amplitude has the opposite sign (lower curve). This
contribution partially cancels the pure Goldstino amplitude, delaying
the onset of the strong coupling regime. For the parameter values used
in Fig.~\ref{fig:partial}, one finds $E_\ast \sim 3.4f$. 

Perturbativity imposes additional constraints, since at high energies
both the Goldstino and the graviton become strongly coupled. These are
of three kinds. First, the interactions in the Volkov-Akulov action
modify $\lambda \lambda \rightarrow \lambda \lambda$ scattering at
one-loop. Compared to the tree-level amplitude, the one-loop amplitude
gives a relative correction of $O(s^2/((4 \pi)^2 f^4))$, which suggests
that the natural cut-off scale is $O(\sqrt{4 \pi} f)$.
At $\sqrt{s} \sim m \sim f$ these corrections are $O(1/(4 \pi)^2)$ and
the expansion parameter is small.

Next, the interactions of the massive graviton provide additional and
stronger constraints.  The ``Einstein gravity'' interactions grow with
energy and become strong at energies of order $E \sim 4 \pi m_{\rm
P}$. With $m_{\rm P} \gsim f /\sqrt{4 \pi}$ this is of order $\Lambda$
or larger. There is however a stronger constraint, since the spin-two
field is massive. The coupling of the longitudinal component becomes
strong at a lower scale, proportional to $(m_{\rm P} m^4)^{1/5}$
\cite{ArkaniHamed:2002sp}. This estimate is obtained from using the
equivalence theorem in the limit $E \gg m$. Factors of $4 \pi$ can be
estimated using naive dimensional analysis \cite{Manohar:1983md}.  For
example, the one-loop contribution of the Goldstino to the vacuum
polarization of the spin-two state becomes comparable to the tree-level
propagator at roughly an energy scale
\begin{eqnarray} 
\Lambda^{(5)}_\ast = (4\pi m_{\rm P} m^4)^{1/5}.
\end{eqnarray} 
One then expects higher dimension operators involving the spin-two field
to be suppressed by this scale.
We will use this scale as a crude estimate of the energy at which the
low-energy effective theory is no longer calculable. Note that for
$m=m_{\rm P}$ the theory becomes strongly coupled at $E \sim 1.7 m$, not
far above threshold.

\begin{figure}[t]
\begin{center}
\includegraphics[width=8cm]{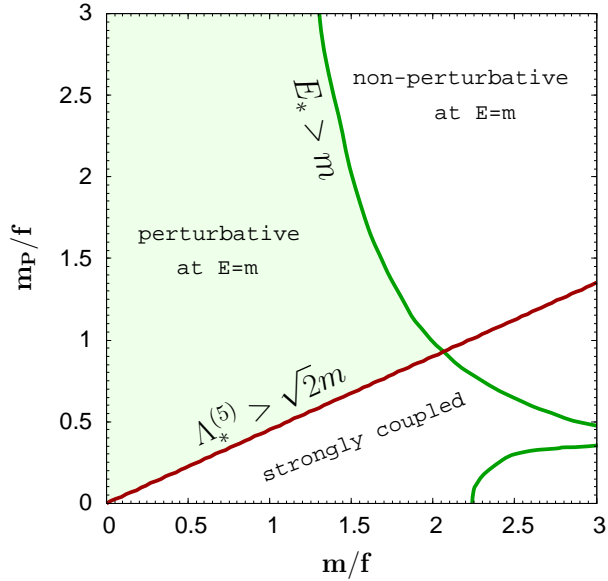}
\end{center}
\caption{The contour of $E_\ast = m$ and $\Lambda^{(5)}_\ast = \sqrt 2
 m$ in the parameter space of $m/f$ and $m_{\rm P}/f$. In the region to
 the left of the thick lines, both $E_\ast > m$ and $\Lambda^{(5)}_\ast
 > \sqrt 2 m$ are satisfied.}  \label{fig:bound}
\end{figure}

The parameter region in which the spin-two resonance can be consistently
incorporated in the effective theory is then approximately bounded by
these considerations.  For illustration, in Fig.~\ref{fig:bound} we plot
the contour of $E_\ast = m$ and $\Lambda^{(5)}_\ast = \sqrt 2 m$ in the
parameter space of $m$ and $m_{\rm P}$.
In the region to the left of the thick lines, both $E_\ast > m$ and
$\Lambda^{(5)}_\ast > \sqrt 2 m$ are satisfied, and we expect that the
production of the spin-two resonance in the single-graviton channel can
be treated in perturbation theory, that is, for $E \sim m < \min [E_*,
\Lambda^{(5)}_*]$.

At higher energies though, the theory becomes strongly coupled. As we
have seen, for generic values of the parameters the scale of strong
coupling is not far above the mass of the spin-two particle.  Since pair
production of the spin-two resonances or scattering of spin-two
resonances requires $E \gsim 2 m$, these processes are generically not
calculable. Thus we have a situation where processes describing the
production and decay of a single on-shell spin-two resonance are plausibly
perturbative, but soon becomes strongly coupled above the
single-particle threshold.

\subsection{Phenomenological Signatures} 

In the effective theory the leading order interactions between the
spin-two and Higgs boson are given by
\begin{eqnarray}
 {\cal K}_{\rm kin.} &= & \delta_{H} \delta^4 (\theta - \lambda)
H^{\mu \nu}
\left[
A_\mu^{\ a} A_\nu^{\ b} (D_a \phi)^\dagger e^{-2 g V} D_b \phi
\right] \nonumber \\ 
& & +  \delta^\prime_{H} \delta^4 (\theta - \lambda)
\tr H 
\left[ \eta^{a b} 
 (D_a \phi)^\dagger e^{-2 g V} D_b \phi
\right]
\label{graviton-Higgs-kinetic term} 
\end{eqnarray}
and
\begin{eqnarray}
 {\cal K}_{\rm pot.} = \delta^{\prime \prime} _H  \delta^4 (\theta - \lambda)
\tr H
\left[
- \delta m^2 \phi^\dagger e^{-2 g V} \phi 
- {\delta k \over 4} \left( \phi^\dagger e^{-2 g V} \phi \right)^2
\right].
\label{eq:traceterm}
\end{eqnarray}
where $\delta_H$, $\delta^\prime_H$ and $\delta^{\prime \prime}_H $ are
parameters assumed to be of $O(1)$.  These interactions begin at
$O(h_{\mu \nu} hh)$ or $O(h_{\mu \nu} VV)$ $(V=W,Z)$. Other interactions
are possible, such as replacing $H_{\mu \nu}$ with $g_{\mu \nu}$. Such
interactions are equivalent to those above, since the difference can be
adsorbed in the normalization of the kinetic and potential terms for the
Higgs.  One may also consider the above interactions multiplied by $\det
A^{-1} \sqrt{-g}$; to zeroth order in $\lambda \partial \lambda$ these
are the same.  The interaction terms proportional to $\tr H$ do not
contribute to any tree amplitude when the spin-two field is on-shell.

A general feature of the spin-two field that distinguishes it from massive
gravitons from extra dimensions (i.e., from a metric), is that it does
not have a minimal coupling. That is, it does not couple universally to
the total stress-energy tensor of matter, or even non-universally to the
stress-energy tensor of each particle. This is explicitly evident in its
couplings to the Higgs boson, Eqs.~(\ref{graviton-Higgs-kinetic term})
and (\ref{eq:traceterm}), since the parameters $\delta_H$,
$\delta^\prime_H$ and $\delta^{\prime \prime}_H $ are unrelated. Only
for a specific ratio of these parameters does the spin-two field couple to
the stress-energy tensor of the Higgs boson.  However in the effective
theory there is no symmetry principle which would enforce such a
condition. 

At one-loop these interactions will generically modify the Fierz-Pauli
form of the graviton mass term. This introduces a ghost at high
energy. For $\delta_H \sim \delta_H^\prime \sim O(1)$, the ghost pole is
above the cut-off scale provided
\begin{eqnarray} 
\Lambda \lsim (4 \pi m^2 m_P )^{1/3} .
\label{ghostpole} 
\end{eqnarray} 
For $m \lsim 4 \pi m_{\rm P}$ the RHS is always larger than the cutoff
$\Lambda^{(5)}_\ast$ above which the graviton is strongly coupled. The
spin-two interactions with the Higgs boson therefore do not introduce a
ghost below $\Lambda^{(5)}_\ast$ provided this condition is satisfied
and the couplings in Eqs. (\ref{graviton-Higgs-kinetic term}) and
(\ref{eq:traceterm}) are no larger than $O(1)$.  It is therefore not
surprising that (\ref{ghostpole}) is also parametrically the same as the
largest possible cutoff of an interacting massive graviton
\cite{ArkaniHamed:2002sp} (i.e., the scale $\Lambda^{(3)}$ in the
notation of that reference). It is perhaps more of a coincidence that
(\ref{ghostpole}) is also the maximum cutoff of an electromagnetically
coupled spin-two particle of charge $e$ \cite{Porrati:2008ha} with the
correspondence $e \rightarrow \delta_H m/m_{\rm P}$.

With the assumption that of the Standard Model particle content only the
Higgs boson couples directly to the strong dynamics breaking SUSY, the
dominant decay modes of the spin-two particle are then
\begin{eqnarray} 
h_{\mu \nu} \rightarrow \lambda {\bar \lambda}, ~hh, ~WW,~ZZ.
\end{eqnarray} 
One obtains 
\begin{eqnarray}
 \Gamma(h_{\mu \nu} \to \lambda \bar \lambda)
= {1 \over 16 \pi}
{1 \over 5}
{m_{\rm P}^2 m^7 \over f^8}
\end{eqnarray}
for the (invisible) Goldstino final state, and 
\begin{eqnarray}
 \Gamma( h_{\mu \nu} \to h h)
= {1 \over 16 \pi}
{1 \over 5}
{\delta_H^2 \over 4}
{m^3 \over 3 m_{\rm P}^2}\left(
1 - {4 m_h^2 \over m^2}
\right)^{5/2}
\end{eqnarray}
for the Higgs boson final state. The powers of velocity appearing here
may be understood by noting that its rest frame, the spin-two particle
couples to the velocity of the Higgs boson, leading to two powers in the
amplitude, or five in the rate.  In the limit $m \gg m_W, m_h$ the
Equivalence Theorem applies, giving $\Gamma(h_{\mu \nu} \to hh) =
\Gamma(h_{\mu \nu} \to ZZ) = \Gamma(h_{\mu \nu} \to WW) / 2$.
Note the spin-two resonance is narrow provided the three mass scales
$m,f,m_{\rm P}$ are all comparable and $\delta^2_H$ is $O(1)$.

We next define a field strength
\begin{eqnarray}
F_{\mu \nu} \equiv {i \over g} [A_\mu^{\ a} D_a, A_\nu^{\ b} D_b],
\end{eqnarray}
by using the ``covariant'' derivative in Eq.~(\ref{eq:cov-der}). A 
graviton-gluon interaction term can then be written as
\begin{eqnarray}
 {\cal K}_{\rm glue} = 
-{\delta_g \over 4k_g} 
\delta^4 (\theta - \lambda)
H^{\mu \rho} G^{\nu \sigma}
\tr \left(
F_{\mu \nu}
F_{\rho \sigma}
\right).
\label{gluon-spin-two interaction}
\end{eqnarray}
The normalization factor $k_g$ is defined by $\tr(T^i T^j) = k_g
\delta^{ij}$. These interactions give 
\begin{eqnarray} 
h_{\mu \nu} \to gg .
\end{eqnarray} 
By assumption the interaction term is small -- $\delta_g \ll 1$ -- and
therefore this decay is suppressed compared to other channels.

This interaction however leads to the production of spin-two particles
through the collisions of gluons.  In the narrow-width approximation the
leading-order differential production cross-section to $hh$ at the
parton level is
\begin{eqnarray}
 \frac{d \hat \sigma}{d \cos \theta} (gg \to h_{\mu \nu}  \to hh)
= 
{1 \over 16 \pi \hat s}
{\delta_g^2 \delta_H^2 \over 2048}
{m^8 \over m_{\rm P}^4}
\left(
1 - {4 m_h^2 \over m^2}
\right)^{5/2}
{\pi \over m \Gamma_{\rm tot}}
\delta (\hat s - m^2)
\sin^4 \theta.
\end{eqnarray}
The phase space is $0 \leq \theta \leq \pi/2$ because of the identical
particles in the final state. For $m \gg m_h$ the two Higgs bosons are
highly boosted. The observation of the $\sin^4 \theta$ dependence of the
cross section will be an interesting confirmation of the spin-two
resonance.

\begin{figure}
\begin{center}
 \includegraphics[height=7cm]{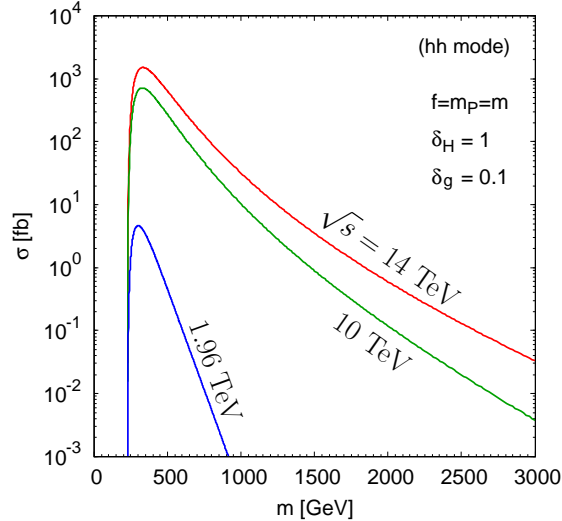}
\end{center}
\caption{The cross section (\ref{total cross-section}) ($hh$ mode) as
 functions of the graviton mass for 1.96, 10 and 14 TeV center-of-mass
 energies. Using CTEQ6M parton distribution functions
 \cite{Kretzer:2003it} and setting $m_h=114$ GeV. The relation $f=m_{\rm
 P}=m$ is kept fixed as $m$ is varied. }  \label{fig:L}
\end{figure}

The total cross section at the LHC is then given by 
\begin{eqnarray}
 \sigma =
\int_0^1 d x_1
\int_0^1 d x_2
\int d \hat s \ \delta (\hat s - x_1 x_2 s)
f_{g} (x_1, m^2)
f_{g} (x_2, m^2)
\tilde \sigma (\hat s) s \delta (\hat s - m^2),
\end{eqnarray}
where
\begin{eqnarray}
 d \hat \sigma = d \tilde \sigma \cdot s \delta (\hat s - m^2)
\end{eqnarray}
and  $s$ is the proton-proton center-of-mass energy.
The expression for the cross-section reduces to
\begin{eqnarray}
 \sigma = {dL (\tau) \over d\tau}  \tilde{\sigma} (m^2)~ ,
 \label{total cross-section} 
\end{eqnarray}
where the luminosity function $dL(\tau)/d\tau$ is defined by
\begin{eqnarray}
 {dL (\tau ) \over d\tau} \equiv
\int_\tau^1 dx\ 
{1 \over x} f_g (x,\tau s) f_g(\tau/x,\tau s), 
\ \ \ \left(\tau \equiv {m^2 \over s} \right) ,
\end{eqnarray}
and the $\tilde \sigma $ factor is
\begin{eqnarray}
 \tilde \sigma (m^2)  =
{\pi \over 64 s} {m^2 \over m_{\rm P}^2} 
\cdot \delta_g^2 \cdot B(h_{\mu \nu} \to hh).
\end{eqnarray}
The cross section (\ref{total cross-section}) for $pp \to h_{\mu \nu}
\to hh$ are shown in Fig.~\ref{fig:L} for a choice of parameters, using
the CTEQ6M parton distribution functions \cite{Kretzer:2003it}. The
spin-two couplings to the Higgs boson and the gluon are chosen to be
$\delta_H=1$ and $\delta_g=0.1$. The production rate depends on the
Higgs boson mass only through phase space, and is therefore significant
only for $2 m_h$ comparable to $m$; in Fig.~\ref{fig:L} we have set
$m_h=114$~GeV.  We have varied $m$ while holding the relations
$f=m=m_{\rm P}$ fixed. The reason for this is to satisfy the
requirements of weak coupling, as discussed previously.  For this choice
of parameters, the cross-section is larger than $1$~fb at 14~TeV
(10~TeV) for spin-two masses $m \lsim 1.75~(1.5)$~TeV. We note that the
production cross-section is sensitive to the spin-two coupling
$\delta_g$ to gluons, so that larger rates are possible.

We conclude this section with comments on other phenomenological
signatures of the spin-two field.

When all the mass scales are comparable and $\delta_H = O(1)$ 
no one decay mode dominates over any other. 
For example, 
with $\delta_H=1$ and $f=m=m_{\rm P} \gg m_h$ one has  
\begin{eqnarray} 
\frac{\Gamma(h_{\mu \nu} \to \lambda {\bar \lambda})}
{\Gamma(h_{\mu \nu} \to hh) 
+\Gamma(h_{\mu \nu} \to ZZ)+\Gamma(h_{\mu \nu} \to WW)} = 3 .
\label{branching ratio} 
\end{eqnarray}
Since producing the graviton in association with a gluon jet is also
proportional to $\delta^2_g$, searching for the invisible decay in this
channel may be promising and important to validate this scenario. Note
monojets are also a generic signature of large extra dimensions
\cite{ArkaniHamed:1998rs}.

Couplings of the spin-two field to electroweak and hypercharge field
strengths may also occur.  These operators are analogous to its
interactions with gluons (\ref{gluon-spin-two interaction}). Although by
assumption they too are suppressed, the rare decay
\begin{eqnarray} 
h_{\mu \nu} \to \gamma \gamma
\end{eqnarray} 
is of obvious experimental interest. The rapidity distribution of the
photons depends on the spin of the resonance, which in principle may be
used to distinguish the graviton from a scalar.

Finally, spin-two particles can also be produced through vector boson
fusion $ q q^\prime \to q q^\prime h_{\mu \nu} $, which has been
recently studied in \cite{vector boson fusion}. In our model the
production rate is proportional to $\delta^2_H$ and therefore under our
assumptions cannot be made arbitrarily small, unlike the production
through gluon-gluon fusion which is suppressed by the small parameter
($\delta _g$).  Compared to Higgs production from vector boson fusion,
the production rate of the spin-two particle is suppressed by a factor
of $v^2/m^2_P$. This is simply because in unitary gauge the amplitude
for producing the spin-two field involves two Higgs vev insertions,
whereas the same amplitude for producing the Higgs boson has only a
single insertion.  The Higgs and graviton production cross-sections
through vector boson fusion are not trivially related however, since
they scale differently with mass due to the dependence on spin. The rate
in this channel could be of experimental interest if the spin-two mass
is low and the scale $m_{\rm P}$ not too large.

Experimentally discovering the spin-two field in different production
channels and measuring the branching ratios to the invisible and all
visible decay channels will obviously help discriminate between
different models of composite or Kaluza-Klein spin-two fields.


\section{Summary}

If SUSY is broken near the TeV scale by strong dynamics then there may
be composites that can be accessed at the LHC. It is desirable to have a
formalism for writing the effective theory describing the interactions
between the matter or gauge fields and the composites, especially in the
situation where the matter and gauge fields are not participants of
these dynamics. The challenge then in this case is that the matter and
gauge fields appear in linearly realized multiplets, whereas the
composites do not.  We have presented a formulation of non-linearly
realized global SUSY in which this can be done.

As an application, we consider two scenarios. In both, the Higgs boson
is such a composite.  No Higgsinos are present in the low-energy theory.
We show that it is possible to write down SUSY invariant Yukawa
couplings and $A$ terms, despite the presence of only one Higgs boson.

Next, we further suppose that the composites include a light spin-two
field in addition to the Higgs boson. The construction of the SUSY
invariant action is in analogy to the Hidden Local Symmetry of chiral
dynamics, where the $\rho$ meson is a massive vector boson of a hidden
local SU(2)$_V$ symmetry. Here though the hidden symmetry is local
Poincar{\' e}. Thus we find that a massive graviton can naturally be
incorporated in a theory with an enlarged spacetime symmetry, which in
this case is global SUSY.

Some phenomenological signatures are discussed. Unlike a generic
Kaluza-Klein graviton, the spin-two particle does not couple to the
stress-energy tensor. Instead its interactions with matter and gauge
fields are constrained only by gauge, Lorentz and non-linear SUSY
invariance. Its dominant decay mode is to Goldstinos (invisible),
electroweak gauge bosons and the Higgs boson. Search strategies to find
boosted Higgs bosons are particularly interesting for this
scenario. Vector boson fusion producing the spin-two particle occurs and
may be of experimental interest. Rare decays to di-photons also occur
but the rate is more model-dependent. Search strategies to find the
Standard Model Higgs boson are therefore simultaneously sensitive to
finding the spin-two particle.

This scenario has the usual low-energy SUSY experimental signatures
(without Higgsinos), while in addition possessing signatures of both
large \cite{ArkaniHamed:1998rs} and warped extra dimensions
\cite{Randall:1999ee} - monojets (ADD) and a spin-two resonance (RS) -
even though there is no extra dimension.  The discovery of SUSY particles and a single
spin-two resonance is not sufficient to claim discovery of an extra
(supersymmetric) dimension. It may just be due to four-dimensional strong SUSY breaking dynamics.

\section{Acknowledgements} 
The work of M.G. and M.K. is supported by the U.S. Department of Energy
at Los Alamos National Laboratory under Contract
No. DE-AC52-06NA25396. Additional support for M.K. is provided by a LANL
Director's Fellowship.  The work of R.K.~is supported in part by the
Grant-in-Aid for Scientific Research (No. 18071001) from the Japan
Ministry of Education, Culture, Sports, Science and Technology.

\end{document}